\documentclass[prd,twocolumn,nofootinbib]{revtex4}
\usepackage{amsmath,amssymb,graphicx,subfigure}

\newcommand{\be}{\begin{equation}}
\newcommand{\ee}{\end{equation}}
\newcommand{\bea}{\begin{eqnarray}}
\newcommand{\eea}{\end{eqnarray}}

\newcommand{\comment}[1]{}

\newcommand{\fd}{\dot{\phi}}

\newcommand{\vf}{\ensuremath{V_{\rm F}}}

\begin{document}
\preprint{hep-th/0603105}
\preprint{UCB-PTH-06/03, LBNL-59843}

\title{Probabilities in the landscape:
  The decay of nearly flat space}

\author{Raphael Bousso and Ben Freivogel}
\email{bousso@lbl.gov, freivogel@berkeley.edu}
\affiliation{Department of Physics and Center for Theoretical
Physics \\
University of California, Berkeley, CA 94720, U.S.A. \\
{\em and}\\
Lawrence Berkeley National Laboratory, Berkeley, CA 94720, U.S.A. }

\author{Matthew Lippert}
\email{lippert@pa.uky.edu}
\affiliation{University of Kentucky, 
Lexington, KY 40506, U.S.A. \\
{\em and} \\
University of Louisville,
Louisville, KY 40292, U.S.A.}

\author{Dedicated to the memory of Andrew Chamblin}


\begin{abstract}
  We discuss aspects of the problem of assigning probabilities in
  eternal inflation.  In particular, we investigate a recent
  suggestion that the lowest energy de~Sitter vacuum in the landscape
  is effectively stable.  The associated proposal for probabilities
  would relegate lower energy vacua to unlikely excursions of a high
  entropy system.  We note that it would also imply that the string
  theory landscape is experimentally ruled out.  However, we
  extensively analyze the structure of the space of Coleman-De Luccia
  solutions, and we present analytic arguments, as well as numerical
  evidence, that the decay rate varies continuously as the false
  vacuum energy goes through zero.  Hence, low-energy de~Sitter vacua
  do not become anomalously stable; negative and zero cosmological
  constant regions cannot be neglected.

\end{abstract}

\maketitle

\section{Introduction}
\label{sec-intro}

In any theory with more than one metastable or stable vacuum, the
question arises with what probability certain low energy physics
phenomena will be observed.  It is not enough to count
vacua~\cite{BP,Dou03,DenDou04b} with the specified property and fold
in anthropic constraints, because the cosmological evolution may favor
some vacua over others.  Whether and how inflationary expansion
factors, decay rates, and initial conditions affect the probability to
be in a given vacuum is a major challenge in theoretical cosmology.
Even once string theory and its vacuum
structure~\cite{BP,KKLT,DasRaj99,GidKac01,DenDou05,DenDou04,ConQue05,%
  BerMay05,BalBer05,SalSil04,SalSil204,MalSil02,Ach02,CarLuk04,%
  CurKra05,GurLuk04,BecBec03,BecCur04,BruAlw04,GukKac03,%
  BucOvr03,DewGir05}, or ``landscape''~\cite{Sus03},
are fully understood, the theory will become predictive only if this
problem is solved.

At the core of the problem is the phenomenon of eternal inflation.  In
a global description, an infinite number of infinitely large regions
containing each vacuum are produced, after unlimited volume expansion
of the preceding vacua.  Elegant proposals to regulate these
infinities have been advanced (see Ref.~\cite{Vil06} for a recent
review, and Ref.~\cite{LinLin94} for an early discussion), but so far
no prescription stands out uniquely, and each has some
counterintuitive properties\footnote{For example, recent
  proposals~\cite{GarSch05,EasLim05}, when applied to a landscape
  consisting of two de~Sitter vacua, would rate them equally likely
  even if their cosmological constants (and thus their entropy) differ
  by an enormous factor.  In an older proposal~\cite{GarVil01}, all
  metastable vacua (including, presumably, our own) would have
  vanishing probability if any stable vacuum exists~\cite{GarSch05}.}.

Proposals for probabilities, and indeed the phenomenon of eternal
inflation itself, are usually formulated in terms of a global,
semiclassical geometry.  In a separate publication~\cite{BouYan06}, it
will be argued that such a description does not exist.  Independently
of this point, however, one has reason to be sceptical of any
simultaneous description of regions that are forever causally
separated.  In the context of black hole formation and evaporation, it
is impossible to reconcile a global description with quantum
mechanical unitarity and linearity~\cite{SusTho93}.  A good theory
should be capable of describing any one observer's experience, for
example an observer inside the black hole, or an observer remaining
forever outside.  A description of two causally disconnected observers
at once, however, does not correspond to any feasible experiment, and
if attempted, leads to contradictions.  Adaptations of this lesson to
the cosmological horizon of de~Sitter space include
Refs.~\cite{CEB2,Fis00b,Ban00,Bou00a,Bou00b,BanFis01a,DysKle02,Sus03}.

This suggests that eternal inflation should be reformulated in terms
of a single causally connected region, or {\em causal
  diamond}~\cite{Bou00a}.  (A causal diamond is the overlap of the
causal past of $q$ with the causal future of $p$, where $p$ and $q$
are any two points on a worldline.  The maximal area on the boundary
of a causal diamond is an upper bound on the entropy in the region it
encloses~\cite{CEB1,CEB2}.)  Perhaps a compelling definition of
probabilities can be given in this observer-centered language.

This program is straightforward to implement for a potential landscape
whose lowest point has positive vacuum energy.  Because the entire
system has finite entropy, it will probe its whole phase space over
and over, and the relative probabilities are easy to obtain.  Under
unitary ergodic evolution, the relative amount of time spent in a
given vacuum, and thus its relative probability, is proportional to
the number of states it contains, ${\cal N}_i$.  (We neglect anthropic
factors in this discussion.)

The entropy a de~Sitter vacuum is given by the area of its
cosmological horizon: ${\cal S}=A_i/4$, where $A_i=12\pi/\Lambda_i$.
Hence,
\begin{equation}
{\cal N}_i = \exp(3\pi/\Lambda_i)
\end{equation}
for a vacuum with cosmological constant $\Lambda_i$.  The lowest
vacuum has the largest entropy and can be thought of as a highly
degenerate ground state for the landscape.  Its exitations correspond
to less entropic configurations, such as a nonempty de~Sitter space.
The other vacua are also merely particular excitations of the lowest
vacuum.

While a landscape with only positive energy vacua is pleasantly
simple, it is also experimentally ruled out.  Dyson, Kleban, and
Susskind~\cite{DysKle02} showed on statistical grounds that our universe
cannot have arisen from the unitary, ergodic evolution of a completely
stable de~Sitter spacetime.  Their analysis can be adapted to any
landscape whose lowest point has positive vacuum energy, with the same
conclusion.  Unless one of the assumptions (unitarity, ergodicity)
breaks down, this implies that the true landscape contains vacua with
nonpositive cosmological constant.

Indeed, the string theory landscape is expected to contain valleys
with negative cosmological constant as well as supersymmetric regions
with vanishing vacuum energy~\cite{KKLT}.  The latter pose a
significant challenge for a causal definition of probabilites, in that
they tend to attract all of the probability, at least according to
very simple probability measures one might propose.\footnote{For the
  same reason, they also promise to admit precise physical
  observables~\cite{FreSus04,Ban04,Bou05,BouFre05}.}  (This is
reminiscent of the difficulties with the global prescription of
Ref.~\cite{Vil98c}, where the probability has support only on
nonpositive vacua.)

To see this, consider two very simple proposals.  First, let us
suppose that the probability for a given vacuum is still proportional
to the amount of time one can spend in it.  This probability measure
has support only for vanishing cosmological constant.  This is because
all the de~Sitter vacua decay after a finite time, and all the vacua
with negative cosmological constant suffer collapse into a big crunch
after a finite time of order $|\Lambda_i|^{-1/2}$.  The $\Lambda=0$
regions, however, are open FRW universes that live forever, so they
dominate.

Second, suppose that the probability of each vacuum is set by the
maximum entropy.  More precisely, suppose it is proportional to the
number of quantum states that can exist within a causal diamond
contained in a corresponding region.  The maximum entropy is set by
the largest area on the boundary of the causal diamond~\cite{Bou00a},
which goes like an inverse power of the cosmological constant.  In
particular, one finds that this area is finite for de~Sitter vacua
(because of the cosmological horizon), as well as for negative energy
vacua (because of their finite lifetime).  On the other hand, the
causal diamonds in open FRW regions with vanishing cosmological
constant can become arbitrarily large, so they dominate.

This would seem to suggest that a more refined probability measure is
required.  Banks and Johnson~\cite{BJ} have recently proposed a
different approach, in which the description of the full string theory
landscape would be quite similar to that of the above toy landscape
with only positive energy vacua.  The lowest {\em positive\/} energy
vacuum would play the role of a degenerate ground state in which the
system spends most of its time.~\footnote{Another discussion of the
  Banks-Johnson proposal can be found in Ref.~\cite{DenDou06}}

In support of this proposal, Banks and Johnson presented numerical and
analytic arguments suggesting that very low energy de~Sitter vacua are
anomalously long-lived.  That is, a vacuum with $\Lambda_{\rm small}$
much smaller than the barrier height would be significantly more
stable than a zero-cosmological constant vacuum obtained by shifting
the entire potential down by the tiny amount $\Lambda_{\rm small}$.
More precisely, the lifetime of the $\Lambda_{\rm small}$ vacuum
would diverge as $\Lambda_{\rm small}$ approaches zero from above, but
would become finite again when $\Lambda_{\rm small}=0$.

We feel that the Banks-Johnson proposal does not remove the
difficulties with $\Lambda=0$ vacua, whose entropy and lifetime can
be arbitrarily large.  Moreover, the proposal involves {\em ad hoc\/}
assumptions that contradict semiclassical results in a regime where
the latter are expected to be valid.  This includes the unavailability
of vacua with small negative cosmological constant, and the ability of
negative energy vacua to ``decay back up'' to the putative de~Sitter
ground state so as to establish detailed balance.

But if these obstacles could be overcome, the Banks-Johnson proposal
would in fact rule out the string theory landscape.  (This was not
noted in Ref.~\cite{BJ}.)  This is because the arguments of Dyson,
Kleban, and Susskind~\cite{DysKle02} would then apply, and the
universe we observe would be overwhelmingly unlikely to arise
dynamically.

In any case, the suggestion that low energy de~Sitter vacua are
anomalously stable is of independent interest and warrants careful
investigation.  It is a radical claim in that it conflicts with
locality.  When a bubble of true vacuum appears it is typically of
microphysical size, and one would not expect it to matter whether the
universe has a cosmological horizon out at, say, $10^4$ Mpc.  Yet,
according to Ref.~\cite{BJ}, it matters a lot: the addition of a small
amount of vacuum energy (the smaller, the better!)  would drastically
suppress the decay of flat space.\footnote{Banks and Johnson suggested
  a holographic explanation of the alleged discontinuity.  For small
  positive $\Lambda$, the false vacuum has enormous entropy due to the
  cosmological horizon.  The argument is that the decay is {\em
    entropically\/} suppressed, because all the horizon entropy would
  be destroyed by the decay.  No such contribution to the entropy is
  present at $\Lambda=0$, so there should be no entropic suppression
  of the transition to a lower vacuum.  However, it is not clear to us
  how adding entropy should stabilize a system.  The cosmological
  horizon is far away when the bubble first forms, and will only be
  destroyed once the bubble has expanded far enough.  Indeed, the same
  logic would imply that the decay of flat space could be suppressed
  by adding enormous entropy to faraway regions (which can be done by
  adding a huge black hole, or with arbitrarily small backreaction
  using radiation).}

In this paper we find that the decay rate is continuous as the false vacuum
energy passes through zero.\footnote{This was discovered independently by A. Aguirre, T. Banks, and M. Johnson, whose results appear simultaneously with this paper.}  Hence the low energy vacua are not
anomalously long lived, i.e., not much longer lived than the flat
space vacua they would be shifted into.

Depending on the detailed shape of the potential, a $\Lambda=0$ vacuum
can be completely stable.  Then continuity demands that the decay of a
false de~Sitter vacuum related by a shift of the potential become
arbitrarily suppressed as the false vacuum energy approaches zero.
(The numerical data presented in Ref.~\cite{BJ} actually pertain to
this case and thus did not support the conclusions drawn there.)
Thus, our results do not exclude the possibility of extremely
long-lived de~Sitter vacua (up to $\log t_{\rm
  decay}\sim\Lambda^{-1}$).  What we rule out is that they will
generically be far more stable than the flat space vacua obtained by
shifting the potential down.

It may be that the $\Lambda=0$ vacua in the landscape are necessarily
supersymmetric.  (Why else would the contributions to the cosmological
constant cancel precisely?)  Then they would all be stable.  However,
this does not automatically imply that the low energy de~Sitter vacua
have lifetimes comparable to the recurrence time, since it is not
clear in which sense those vacua are close to supersymmetric.  In the
real landscape, we do not get to shift the potential continuously.  A
low energy de~Sitter vacuum may be very far from a supersymmetric
region in field space.

The observation of Banks and Johnson stands that the decay of
de~Sitter space is exponentially suppressed while negative energy
regions meet their demise in polynomial time.  The potential
significance of this asymmetry remains to be explored.  It would be
premature to conclude, however, that negative energy vacua can simply
be neglected.  In any case, a more immediate challenge is that the
$\Lambda=0$ vacua appear to dominate, both in terms of entropy and in
terms of their lifetime.  This is problematic because we do not appear
to live in one.

This paper is structured as follows.  In Sec.~\ref{sec-cdl} we review
the instanton calculus describing the decay of a false vacuum in the
presence of gravity.  In Sec.~\ref{sec-cont}, we demonstrate the
continuity of the decay rate.  The key ingredient in our argument is
to show and exploit that singular solutions nearby the regular
instantons behave continuously as the false vacuum energy is taken to
zero, unlike the regular instanton geometry itself.

In Sec.~\ref{sec-structure} we give a thorough analysis of the space
of solutions to the Coleman-De Luccia equations, depending on the
initial value of the field, and on the energy of the false vacuum.  We
now also explore regions where the latter is far from zero.  In
Sec.~\ref{sec-numerical} we present numerical evidence supporting our
analytical results.

\section{Coleman-De Luccia tunneling}
\label{sec-cdl}

In this section, we briefly review the decay of a false vacuum,
following the analysis of Coleman and De Luccia~\cite{CDL}.  (For
earlier work that neglects effects of gravity, see
Refs.~\cite{Col77,CalCol77}.)  We set up notation, mostly following
Banks and Johnson~\cite{BJ}.

Consider a scalar field $\phi$ with potential $V(\phi)$.
We assume that $V$ has two local minima, at $\phi_{\rm F}$ and
$\phi_{\rm T}$, with $V(\phi_{\rm F})>V(\phi_{\rm T})$.  Examples are
shown in Fig.~\ref{fig-potential1} and Fig.~\ref{fig-potential3}
below.  Without loss of generality, we take the top of the barrier to
be at $\phi=0$ and the true vacuum at $\phi_{\rm T}>0$.

Let us assume that initially the field is in the false vacuum
thoughout space: $\phi=\phi_{\rm F}$.  Classically, it would remain
there forever.  Quantum mechanically, it may be possible for it to
lower its energy by tunneling through the barrier towards the true
vacuum.  The path integral for this process can be approximated by a
regular Euclidean solution, or instanton.

The most symmetric (and, presumably, dominant) nontrivial Euclidean
solution is $SO(4)$ invariant, with metric
\begin{equation}
ds^2 = dt^2 + \rho(t)^2 d\Omega_3^2~,
\end{equation}
Here $d\Omega_3^2$ is the metric on the unit three-sphere; we will
think of the parameter $t$ as Euclidean time.  Thus the instanton is
described by two functions $\phi(t)$ and $\rho(t)$.  

\begin{figure}[!htb]
\subfigure[]{
\includegraphics [scale=.3]
{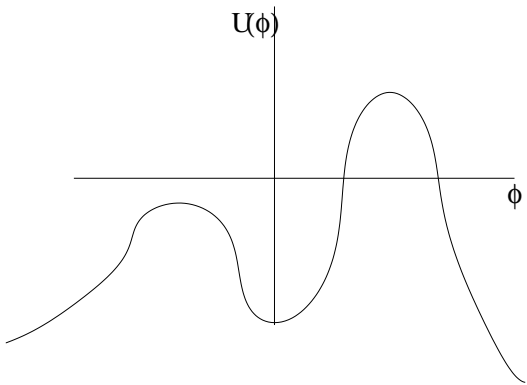} }
\hspace{0.3 in}
\subfigure []{\includegraphics [scale = .3]{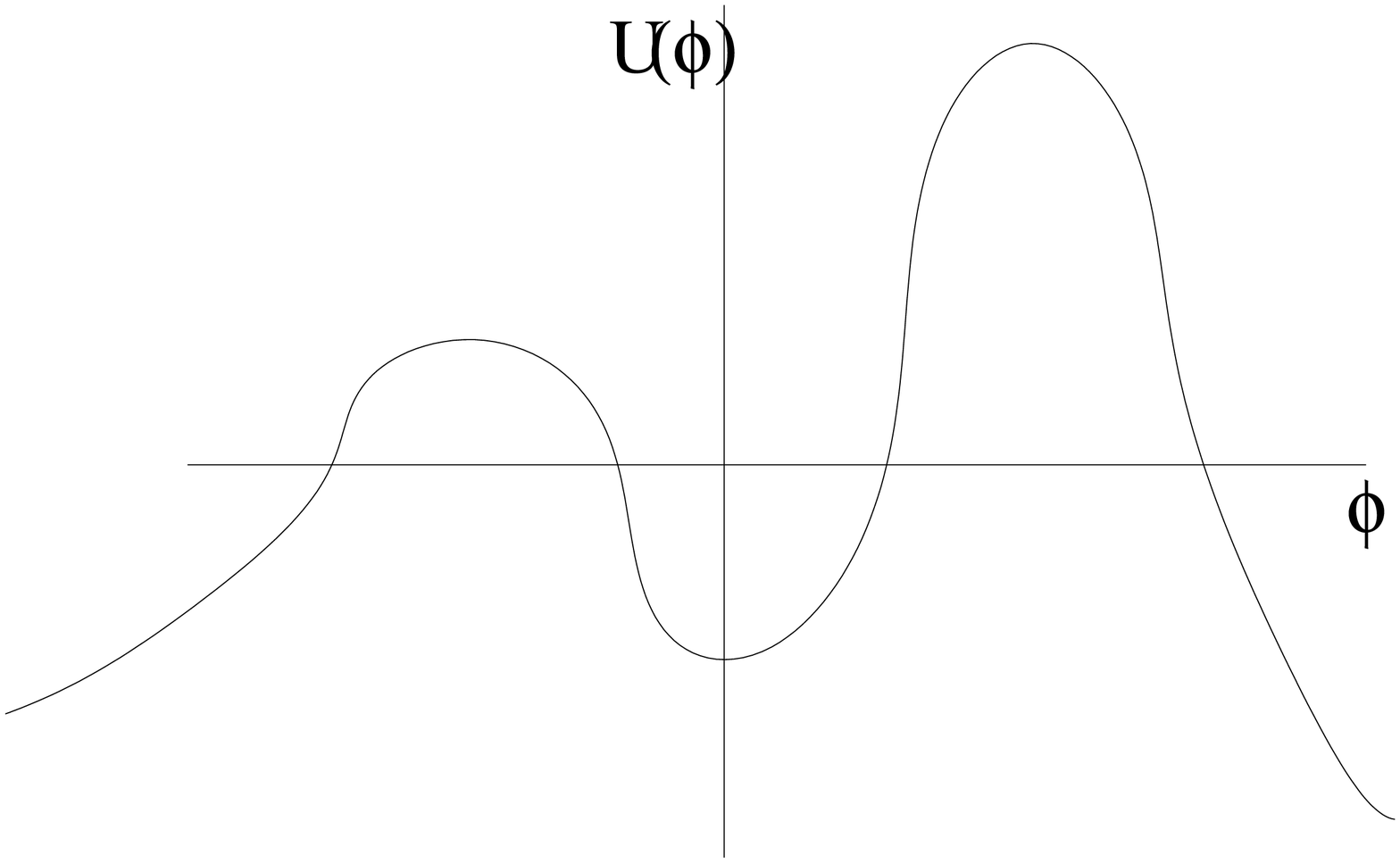}}
\caption{The Euclidean potentials we are interested range from that
  shown in (a) to that in (b).}
\label{fig-eucpot}
\end{figure}

The equations of motion are Euclidean versions of the FRW equations
for a closed universe with a scalar field:
\begin{eqnarray} 
\label{unscaledCdL}
&&\ddot{\rho} = -\frac{4 \pi}{3 M_{\rm Pl}^{2} }  \rho \left[\dot{\phi}^2 + V(\phi) \right]~, \\
&&\ddot{\phi} + 3 \frac{\dot{\rho}}{\rho} \dot{\phi} = V'(\phi)~,
\end{eqnarray}
where an overdot (prime) denotes differentiation with respect to $t$
($\phi$).  They obey the constraint
\begin{equation}
\label{eq-constr}
\dot{\rho}^2-1 = \frac{8 \pi}{3 M_{\rm Pl}^{2}} \rho^2 \left[\frac{\dot{\phi}^2}{2}-V(\phi)\right]~.
\end{equation}
Thus, the scalar field behaves like a particle moving in the Euclidean
potential $U=-V$ (see Fig.~\ref{fig-eucpot}), with friction
proportional to $\dot{\rho}/\rho$.

There are compact and noncompact solutions.  Noncompact solutions
have one pole (a value of $t$, conventionally taken to be zero, at
which $\rho$ vanishes), and have the topology of $\mathbf{R}^4$.
Compact instantons have two poles and have the topology of a
four-sphere.  For compact solutions, $\dot{\rho}$, and thus the
friction, eventually becomes negative.

At any pole, continuity of the field gradient requires
\begin{equation}
\dot{\phi} = 0~~~~~(\mbox{at}~\rho=0)~.
\end{equation}
Moreover, the absence of a conical singularity requires the
boundary condition
\begin{equation}
|\dot{\rho}|=1~~~~~(\mbox{at}~\rho=0)~.
\label{eq-rdot}
\end{equation}
At the first pole, from which the equations are integrated, this is
imposed as a boundary condition.  For compact solutions, the
constraint equation~(\ref{eq-constr}) guarantees that
Eq.~(\ref{eq-rdot}) will automatically be satisfied at the second
pole, as long as it has been ensured that the scalar field energy
\begin{equation}
E = \frac{\dot{\phi}^2}{2}-V(\phi)
\end{equation}
remains bounded.  For compact solutions, Eq.~(\ref{eq-rdot}) implies
divergent antifriction at the far pole.

Let us first consider the case where the false vacuum has zero or
negative energy: $V(\phi_{\rm F})\leq 0$.  Decay is mediated by an
instanton that is asymptotic to the background, i.e., to Euclidean
flat or AdS space.  Hence, the instanton must be noncompact, with
$\phi \to \phi_{\rm F}$ for $\rho \to\infty$.  It must contain a bubble of true
vacuum: $\phi>0$ at $\rho=0$.  As we shall discuss in more detail below, it
is by no means automatic that such an instanton exists.  If it does
not, the false vacuum is stable.  If it does, then the rate of decay
per unit volume is given by
\begin{equation}
\Gamma\sim\exp[-(S_{\rm inst}-S_{\rm bg})]~.
\label{eq-rate}
\end{equation}
Here, $S_{\rm inst}$ is the Euclidean action of the instanton, and
$S_{\rm bg}$ is the action of the background solution.  (This is the
solution corresponding to $\phi(t)=\phi_{\rm F}$ for all $t$, i.e.,
Euclidean flat or AdS space.)  

The action is given by
\begin{equation}
S = \int d^4x \sqrt{g} \left[ -\frac{M_{\rm Pl}^2}{16\pi}\ R +
  \frac{1}{2} (\nabla\phi)^2 +V(\phi) \right]~.
\end{equation}
On shell, this becomes simply
\begin{equation}
S = -\int d^4x \sqrt{g} V(\phi)~.
\end{equation}


The energy obeys
\begin{equation}
  \dot{E} = -3\frac{\dot{\rho}}{\rho}\dot{\phi}^2~.
\end{equation}
Since $\dot{\rho}>0$ for noncompact solutions, it is clear that energy
will be lost to friction.  Hence, the scalar must start out, at $\rho=0$,
at some value $\phi_0$ where the Euclidean potential is higher than in the true
vacuum: $U(\phi_0)>U(\phi_{\rm T})$.  A suitable point may or may not exist.
By the same token, it is clear that there is never an instanton
describing the formation of bubbles of false vacuum in a true vacuum
with nonpositive energy.

Now let us turn to the case where the false vacuum is de~Sitter space:
$V(\phi_{\rm F})>0$.  (The above equations for the action and the
energy still stand.)  In this case the background action in
Eq.~(\ref{eq-rate}) is finite, and tunneling is always allowed.  (If
all else fails, the Hawking-Moss instanton, $\phi \equiv 0$, describes
tunneling to the top of the barrier~\cite{HawMos83}; see also
Ref.~\cite{Lin98}.)  However, if the background action dominates over
the instanton action, tunneling will be extremely suppressed:
$\log\Gamma\sim\exp S_{\rm bg}$, corresponding to a lifetime of order the
Poincar\'e recurrence time in the background de~Sitter space.

The instanton, in this case, will be compact and will not reach either
$\phi_{\rm T}$ or $\phi_{\rm F}$ exactly.  As we will discuss in more
detail in Sec.~\ref{sec-structure}, there can be several such
instantons.  We will be most interested in one-pass instantons, which
cross the barrier precisely once.  An example (type I) is an instanton
that resembles the de~Sitter four-sphere background solution except in
a small region containing a bubble of true vacuum.  Then the action
difference results from the bubble region alone, and the rate will not
depend strongly on the background cosmological constant.  Another
example (type II) is an instanton that does not spend much time near
$\phi_{\rm F}$.  Then the background action dominates, and tunneling is
suppressed by the inverse background cosmological constant in the
exponent.\footnote{If the both vacua have positive energy, then
  tunneling is possible in both directions, mediated by the same
  instanton.  The difference in rates comes entirely from the
  different background actions that must be subtracted depending on
  the direction of the process.  From Eq.~(\ref{eq-rate}) one finds
  that $\Gamma_\uparrow/\Gamma_\downarrow\sim\exp(S_{\rm T}-S_{\rm
    F})$.  The action of Euclidan de~Sitter space is minus the entropy
  ($S=- {\cal S}$).  Thus, this result agrees with statistical
  expectations, as discussed in the introduction.  Unless the two
  vacuum energies are very similar, the upward decay is an example of
  a highly suppressed process where the background action dominates.}

In the following we will investigate how the tunneling rate depends on
overall shifts of the potential.  We write
\begin{equation}
V(\phi) = V_0(\phi)+V_{\rm F}~,
\end{equation}
where $V_0(\phi_{\rm F})=0$.  The additive constant $V_{\rm F}$ shifts
the entire potential up or down, with $V_{\rm F}$ being the false
vacuum energy.

Although we will explore the parameter space quite broadly, we are
interested mainly in confirming the continuity of the rate as the
false vacuum energy passes through zero.  For potentials $V$ that
permit the decay of flat space, we will find that the addition of a
small cosmological constant $V_{\rm F}$ does not change the rate much,
leading to a type I decay.  If flat space is stable, then the addition
of a small cosmological constant leads to a highly suppressed, type II
decay of the resulting de~Sitter space.  These results express the
continuity of the decay rate.

In the next section we develop analytical arguments in support of
these assertions.  Global properties of the solution space will be
discussed in Sec.~\ref{sec-structure}.  The final section contains
numerical evidence.

\section{Continuity of the decay rate of nearly flat space}
\label{sec-cont}

In this section, we present our argument that the decay rate is
continuous as the potential is shifted by a constant and the false
vacuum energy approaches zero from above ($V_{\rm F}=0$).  Our argument
combines a number of intermediate results, which will also be useful
in later sections.  Hence we will structure this section by developing
individual results in separate subsections.

We will rely heavily on the properties of singular solutions.  {\em
  Hence, from here on, when we say ``solutions'', this includes
  singular solutions, i.e., integrals of the equations of motion that
  run into a singularity.  We will write explicitly ``singular
  solutions'' or ``regular solutions'' to refer specifically to one of
  these classes.  ``Instanton'' means ``regular solution''.  When we
  say ``compact'', we mean that the radius approaches a second zero,
  independently of whether it does so in a singular or regular way.
  Throughout the paper, we use the term ``generic'' in the technical
  sense: namely, a feature is generic if it occurs in an open set of
  moduli space. In other words, generic means ``unchanged by
  infinitesimal perturbations."}

\subsection{Solutions are generically compact, and all singular
 solutions are compact} 
 \label{subsec-com}

It is important to our method that solutions are generically compact,
so we will give a careful argument.  By construction, all of our solutions
start with regular initial conditions at $\rho = 0$; then we
evolve in Euclidean time to find the full solution. We claim that
generically $\rho$ returns to $0$; in particular, it will do so for
all singular solutions.

For the sake of this argument, we assume that outside the region of
interest the Euclidean potential is negative everywhere and has no
extrema, and also that it stays finite everywhere. Recall that we are
assuming that the local minimum in the Euclidean potential is
negative, and that the potential has no other extrema.

There are three ways one might imagine that the radius $\rho$ can fail
to return to zero:

(1) $\rho$ could remain finite and the solution could remain smooth 
as $ t \to \infty$.

(2) Evolution could stop at a singularity at finite $\rho$.

(3) $\rho$ could asymptotically approach infinity as $t \to \infty$.

(4) $\rho$ could diverge at finite $t$.

We will see that possibilities (1), (2), and (4) are forbidden by the
equations of motion, while possibility (3) occurs nongenerically. An
example of option (3) is the usual instanton which mediates the decay
of a false non-de~Sitter vacuum.

To see that possibility (1) is forbidden, we focus on the evolution of
the Hubble parameter, $H = \dot{\rho}/\rho$. Combining the FRW equations
(\ref{unscaledCdL}) gives
\begin{equation} 
  \dot{H} = - 4 \pi M_{\rm Pl}^{-2} \fd^2 - 1/\rho^2~.
\label{eq-hdot}
\end{equation} 
The right side is negative semidefinite, so $H$ can only decrease.  

We argue by contradiction. Suppose there exists a maximum radius $\rho_{\rm max}$.  Then $\dot{H}$ is bounded above by $-1/\rho_{\rm max}^2$.  This implies that $H \to -\infty$ as $t \to
\infty$. Now 
\begin{equation}
  \rho(t) = \int_0^t{dt' \dot{\rho}(t')} = \int_0^t{dt' \rho(t') H(t')} 
\end{equation}
By assumption, $\rho(t)$ stays finite, and it must be positive because
it is the radius of the $3$-spheres. Since we have just shown that $H$
diverges to negative infinity and by assumption $\rho(t)$ is finite
and positive, the integral diverges. This contradicts our assumption
that $\rho$ remains finite.

Possibility (2) is a singularity at finite $\rho$. To achieve a
singularity at finite $\rho$, we would need $\dot{\rho}$ and/or $\fd$
to diverge.  We choose the singularity to occur at $t=0$ and
characterize the divergent behavior of $\dot{\rho}$ and $\dot{\phi}$
there. In order for $\dot{\rho}$ to diverge without $\rho$ diverging,
we need $\rho \sim t^p$ for $0 > p > -1$. The FRW equation is
\begin{equation}
  \dot{\rho}^2-1 = \frac{8 \pi}{6 M_{\rm Pl}^{2}}\rho^2 
  \left[\frac{\dot{\phi}^2}{2}-V(\phi)\right]~.
\end{equation}
By assumption, $V$ is finite everywhere and the singularity forms at
finite $\rho$.  Keeping just the singular terms the equation becomes
\begin{equation}
  \dot{\rho}^2 = \frac{8 \pi}{6 M_{\rm Pl}^{2}} \rho^2 \dot{\phi}^2~,
\end{equation}
implying that $\fd$ diverges in the same way as $\dot{\rho}$, $\fd
\sim t^p$.  Now the equation of motion for $\phi$, keeping divergent
terms, is
\begin{equation}
\ddot{\phi} + 3 \frac{\dot{\rho}}{\rho} \dot{\phi} = 0~.
\end{equation}
With our assumptions so far, the first term diverges as $t^{p-1}$,
while the second term diverges as $t^{2p}$. Since we require $0 > p >
-1$, the two terms diverge as different powers of $t$ and the equation
cannot be solved.

Possibility (4) is forbidden because
\begin{equation}
\rho(t) = \rho(t_0) \exp \left( \int_{t_0}^t H(t') dt' \right) ~.
\end{equation}
By Eq.~(\ref{eq-hdot}), H is monotonically decreasing.  Thus, if
$\rho$ and $H$ are finite at some time $t_0$, then $\rho$ cannot
diverge in finite time.

This leaves only possibility (3), a regular solution in which $\rho \to
\infty$.  This is not forbidden, but as we will now show, it is
nongeneric, as it requires infinite fine-tuning of the initial
conditions. The basic idea is that if the radius $\rho$ is to approach
infinity, then the field will have to asymptotically approach one of
the vacua so as to provide a resting place with negative cosmological
constant. Since the field must come to rest at a local maximum of the
Euclidean potential, this requires infinite tuning.

To show that the field must approach a vacuum, let us recall that
$\dot{H}$ is negative semidefinite and $H$ must remain nonnegative for
a noncompact solution.  Hence, $\dot{H}$ must approach zero as $\rho
\to \infty$. Further, Eq.~(\ref{eq-hdot}) indicates that $\dot{H} \to 0$
requires $\fd \to 0$.  Using this information, the Friedmann equation
Eq.~(\ref{eq-constr}) becomes, as $\rho \to \infty$,
\begin{equation}
H^2 =-\frac{8 \pi}{3 M_{\rm Pl}^{2}}  V(\phi),
\end{equation}
which requires the potential $V$ to be negative. 

Finally, recall that the equation of motion for $\phi$ is
\begin{equation}
\ddot{\phi} + 3 \frac{\dot{\rho}}{\rho}\dot{\phi} = V'(\phi)~.
\end{equation}
We have already established that $\fd \to 0$, so we can ignore the
second term. In order for $\fd$ to remain zero, we need $\ddot{\phi}
\to 0$, so $V'(\phi) \to 0$.  Hence the field approaches a critical
point of the potential.  Since we also showed that $V<0$ as
$\rho\to\infty$, this means that $\phi$ asymptotes to one of the
vacua.  (As discussed in the next section, this is necessarily the
false vacuum, though this is not crucial here.)

Now it is easy to argue that this solution is nongeneric.  Approaching
the top of a hill as $\rho \to \infty$ is highly unstable. An
infinitesimal perturbation of the starting point will cause the field
to overshoot or undershoot the top of the hill, leading to a compact
singular solution.

This completes our argument that compact solutions are generic.  We
have also seen that noncompact solutions are necessarily regular.
Hence, all singular solutions are compact.

\subsection{Noncompact regular solutions are one-pass}
\label{noncom}

Noncompact solutions have no anti-friction; as shown above, $H \ge 0$
throughout the solution. Since there is no anti-friction, the energy
can only decrease. If the field starts near the false vacuum, it has
no hope of achieving the true vacuum, which is at a higher Euclidean
energy. If it starts near the true vacuum, it cannot have a turning
point because a turning point requires $E < U_{\rm F} < U_{\rm T} $, so the field
will never have enough energy to ascend to the top of either vacuum.

To summarize, the only possible noncompact solution is a one-pass,
regular solution with the field near the true vacuum at the origin and
asymptoting to the true vacuum at infinity.

For oddly shaped potentials, there may even be more than one such
solution.  Such potentials can retain their odd properties under small
deformations, so using our technical definition we cannot call them
nongeneric.

\subsection{The field escapes to $\pm\infty$ at the singularity}
\label{sec-escape}

We showed above that all singular solutions are compact, with the
singularity developing as the second zero of $\rho$ is approached.  It
follows that the behavior of the fields near the singularity is
universal.  In particular, the field $\phi$ escapes in the direction
indicated by the values of $\fd$ sufficiently close to the second
pole.

To show this, recall that the potential is finite everywhere. We write
the equations of motion, keeping only terms which will diverge near
the pole.
\begin{eqnarray}
\ddot{\phi} + 3 H \dot{\phi}  &=& 0 \\
H^2 = 4 \pi M_{\rm Pl}^{-2} \dot{\phi}^2 &+& 1/\rho^2
\end{eqnarray}
To find the divergent behavior, we assume that $\dot{\phi}$ and $\rho$
are both power-law in $t$. We find
\begin{equation}
a(t) = A t^{1/3} \ \ \ \ \ \ \ \fd (t) = B/t
\label{eq-div}
\end{equation}
where we have taken the pole to be at $t=0$. Note that $\fd$ remains
of constant sign, as advertised.  The field value $\phi$ diverges
logarithmically.

\subsection{Across a regular compact solution, the number of passes
  generically changes by one.}
\label{sec-jump}

Consider perturbing the starting point of a compact instanton.  If the
perturbation is very small, then the new solution is virtually
unchanged for a long time.  But we know that near the opposite pole
the field must blow up.  This is because the new solution will fail to
approach the pole with just the right conditions to achieve $\fd = 0$
at the pole, and diverging anti-friction will magnify the mistake.

Generically, perturbing the starting point in opposite directions will
result in opposite runaway of the field at the far pole.  A
perturbation in one direction will result in $\fd$ reaching zero
already for finite (small) $\rho$.  Then the field changes direction,
and the singular behavior of Eq.~(\ref{eq-div}) will push it off to
infinity.  Hence, it will have one additional pass compared to the
regular solution.

Under a perturbation of the starting point in the other direction,
$\fd$ will not reach the zero that it reached for the unperturbed
instanton at the far pole.  The divergence will push it off to infinity
with no additional pass compared to the regular solution.  Hence, the
number of passes changes by one across a regular compact solutions.

Things are more complicated for noncompact instantons.  Because there
is no anti-friction, we cannot appeal to the universal divergent
behavior of Eq.~(\ref{eq-div}).  Hence, the details of the potential
can be important. There is no general rule about how the number of
oscillations changes across a noncompact instanton.  Indeed, we will
provide analytic and numerical evidence that jumps can be by more than
one oscillation.

Going in the other direction, we expect that if two nearby starting points 
$\phi_1$ and $\phi_2$ result in singular solutions with different number of passes
$p_1$ and $p_2$, there will be at least one regular solution with
starting point between $\phi_1$ and $\phi_2$.  However, we do not
prove this rigorously.

\subsection{The decay rate is continuous near $V_{\rm F}=0$}
\label{subsec-cont}

What makes it hard to rule out a discontinuity in the tunneling rate
at $V_{\rm F}= 0$ is the fact that the regular instanton {\em does\/}
change form: for $V_{\rm F}>0$ it is compact, while for $V_{\rm F} =0$
it is noncompact. As a result, doing perturbation theory in $V_{\rm
  F}$ around the point $V_{\rm F}= 0$ is confusing.  We circumvent
this problem by perturbing the {\em singular\/} solutions near the
instanton.

There are two cases.  We defer the case where flat space is stable,
and begin with the more interesting case where it can decay.  Thus, we
assume that there exists a noncompact instanton mediating the decay at
$V_{\rm F} =0$.  We want to prove the existence of an instanton for
infinitesimal positive $V_{\rm F}$ whose action scales with the de Sitter
radius in the false vacuum.

Consider perturbing the starting point of the $V_{\rm F} = 0$
instanton by a small amount $\delta\phi$. The instanton itself is a
solution in which the field starts near the true vacuum at $\rho = 0$ and
approaches the false vacuum as $\rho \to \infty$. Generically, perturbing
the starting point in one direction will cause the field to overshoot
the false vacuum, leading to a singular one-pass solution, while
perturbing the starting point in the other direction will cause the
field to undershoot the false vacuum, resulting in a singular solution
with at least two passes.\footnote{One might imagine a situation where
  perturbing the starting point in either direction has the same
  effect, say undershoot. This is conceivable, but it is requires the
  first-order change to accidentally be zero at the instanton, which
  is nongeneric.} The instanton is at the boundary between the
singular one-pass overshooting solution and the singular multi-pass
undershooting solution.

So there is the regular $V_{\rm F} = 0$ instanton, with starting point
$\phi_0$, and on either side of it there are singular solutions with
starting points $\phi_0 \pm \delta \phi$.  The regular instanton
arrives at the false vacuum at $t= \infty$.  The overshooting solution
arrives at the false vacuum too soon.  However, the time when it does
goes to infinity as $\delta \phi \to 0$. Similarly, the undershooting
solution reaches its turning point ($\fd=0$) at a time which goes to
infinity as $\delta \phi \to 0$.

Now let us ask what happens to the {\em singular\/} solutions if we
increase $V_{\rm F}$ while leaving the starting point fixed. This is a
perturbation of the potential rather than of the initial value.  If
$V_{\rm F}$ is very small, then the undershooting solution will still
undershoot and the overshooting solution will still overshoot.  By the
results of Sec.~\ref{sec-jump}, between these two singular solutions
must lie at least one regular compact instanton.

Having demonstrated the existence of an instanton, we will now argue
that the associated rate of decay changes continuously as $V_{\rm F}$
is increased away from zero.  We have noted that the time at which the
singular solutions at $V_{\rm F}=0$ hit the singularity can be made
arbitrarily large by choosing the perturbation $\delta\phi$ to be
small. By continuity of the singular solutions and by interpolation,
this implies that a regular instanton exists at positive $V_{\rm F}$
whose size increases without bound as $V_{\rm F} \to 0$.  Continuity
and interpolation also imply that the size of the true vacuum bubble
is roughly the same both for the singular and regular solutions, and
both at $V_{\rm F}=0$ and infinitesimal positive $V_{\rm F}$.  Hence
the $V_{\rm F}>0$ singular solutions will spend all but a fixed amount
of their time near the false vacuum\footnote{In fact, exponentially
  close.} before over- or undershooting.  Therefore, the regular
$V_{\rm F}>0$ instanton will be virtually identical to the background
de~Sitter space over a volume that diverges as $V_{\rm F}\to 0$,
differing only in a ``bubble volume'' that remains finite as $V_{\rm
  F}\to 0$.  Hence, we expect the $V_{\rm F}>0$ decay rate to be
comparable to the $V_{\rm F}=0$ decay rate.


In the case where flat space is stable, continuity simply means that
the decay rate should vanish as $V_{\rm F} \to 0$.  The assumption of
stability requires that any compact regular solution at some
infinitesimal $V_{\rm F}>0$ is deformed into a regular solution that
is still compact, with finite action, at $V_{\rm F}=0$.  Meanwhile,
the background instanton action diverges as $V_{\rm F}\to 0$.  By
Eq.~(\ref{eq-rate}), the decay rate approaches zero.

\section{The CDL solution space}
\label{sec-structure}

\subsection{Classification and diagrams}
\label{sec-a}

In this section, we use the results derived above to develop a full
picture of the Coleman-De Luccia solution space.  We will be
interested in how regular solutions can appear or disappear as the
false vacuum energy $V_{\rm F}$ is varied beyond the infinitesimal
neighborhood of $V_{\rm F}=0$.  We continue to distinguish the two
important cases discussed above: whether tunneling is allowed at
$V_{\rm F}=0$ or not.

\begin{figure}
\begin{center}
\includegraphics[scale = .5]{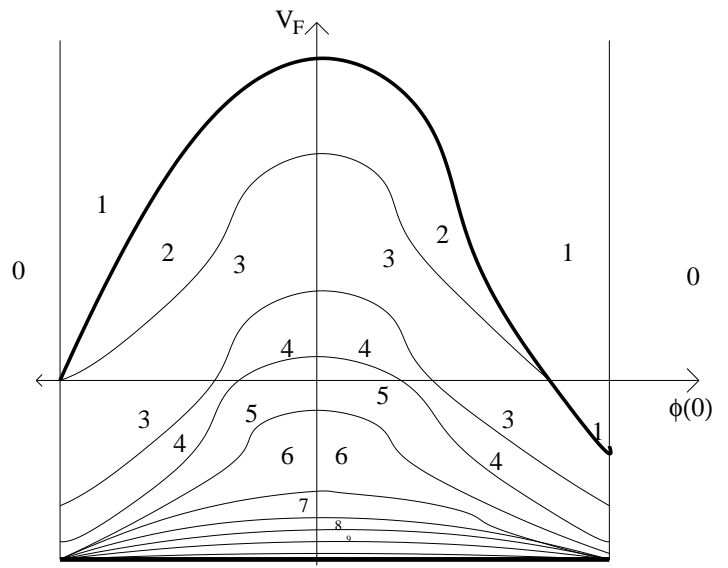}
\end{center}
\caption{A diagram showing solutions with starting point $\phi(0)$ at
  the regular pole, in a potential shifted vertically by $V_{\rm
    F}$. The numbers indicate the number of passes for each solution
  before the field escapes to infinity. The lines divide regions with
  different numbers of passes and represent regular solutions. In this example,  tunneling is allowed for $V_{\rm F} = 0$. The instanton of
  interest is the thick line. For positive $V_{\rm F}$, the instanton is
  compact, and the part of the line at negative $\phi(0)$ represents
  the same solution, thinking of the opposite pole as the starting
  point. For $V_{\rm F} \leq 0$, the instanton becomes noncompact so there is
  only one origin of polar coordinates.} 
\label{fig-tunnel}
\end{figure}

\begin{figure}
\begin{center}
\includegraphics[scale = .5]{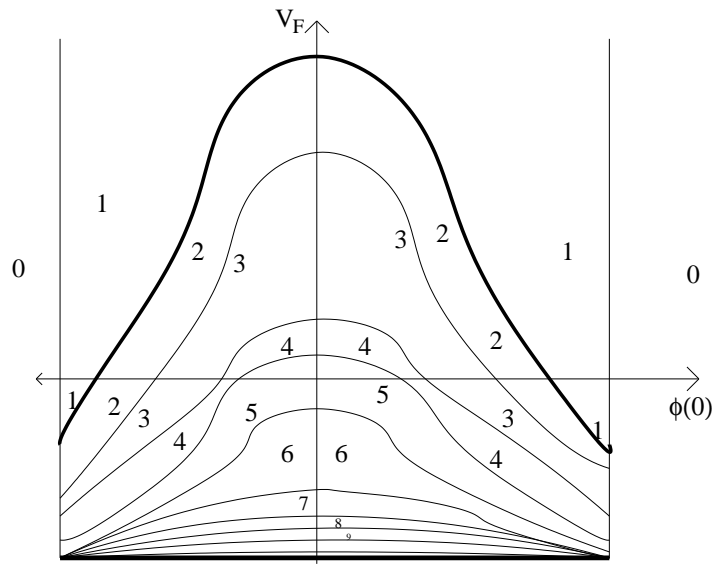}
\end{center}
\caption{Here, tunneling is forbidden for $V_{\rm F} \leq 0$. The
  single-pass instanton is the thick line; for a given $V_{\rm F}$,
  there are two values of $\phi(0)$ which represent the field values
  at the two poles. For negative $V_{\rm F}$ the instanton remains
  compact, so it no longer mediates decay.}
\label{fig-notunnel}
\end{figure}

For each of these cases, we have picked a simple potential $V_0$.
Figures~\ref{fig-notunnel} and~\ref{fig-tunnel} show solutions as a
function of the starting point $\phi_0$ and the parameter $V_{\rm
  F}$.\footnote{This diagrammatic technique was explained to us by
  Matthew Kleban; see Batra and Kleban,  {\it to appear}.}  
  More complicated diagrams are possible for
different potentials. Our goal here is to analyze what appear to be
the two simplest situations.  Their structure is surprisingly rich.
Here we will analytically explain the features they show; in the
following section, we will confirm them numerically.

Recall that the top of the barrier (the minimum of the Euclidean
potential $U$) has been chosen to reside at $\phi=0$.  As long as
$U(0)<0$, a generic starting point $(V_{\rm F},\phi_0)$ will yield a
singular solution.  We classify each starting point by how many times
the field passes through the local minimum of the Euclidean potential
before escaping to infinity, the {\em number of passes\/}.  These
two-dimensional sets of points will be separated by one-dimensional
lines corresponding to the regular solutions.

Note that in these diagrams a regular compact
solution with an odd number of passes is represented by two points in
the diagram, because either pole could be considered the starting
point.  All other solutions, in particular all singular solutions, are
represented by one point, the value of the field at the regular pole.

The fact that we can define the number of passes on a space consisting
of the starting point and \vf\ , as sketched in the diagrams,
immediately tells us that regular solutions cannot simply disappear as
we dial $V_{\rm F}$; they must annihilate with other regular solutions
in a consistent way. For example, a fairly general phenomenon is the
following.  As $V_{\rm F}$ is increased, raising the potential up to
higher cosmological constant, some instantons disappear, annihilating
at $\phi = 0$. What is happening is that near the bottom of the
Euclidean potential well there is a characteristic frequency of
oscillation which is independent of $V_{\rm F}$. There is also a
characteristic de Sitter time given by the size of the instanton with
the field sitting at the bottom- the Hawking-Moss instanton.
Increasing $V_{\rm F}$ decreases the de Sitter time, so there is no
longer enough time to have as many oscillations. In the figure, one
can see the regions with multiple oscillations disappearing one by one
as $V_{\rm F}$ grows until eventually only the Hawking-Moss instanton
is left (the vertical line at $\phi=0$).

In our figures, the numbers increase monotonically
towards $\phi = 0$ for fixed $V_{\rm F}$ , meaning 
that starting points closer to the minimum
allow more oscillations. This is a common situation, but as described by
Hackworth and Weinberg~\cite{HacWei04}, more interesting possibilities
are not difficult to arrange.

Our method is extremely useful in constraining the presence and
location of regular solutions, and it generalizes to a variety of
situations. For example, one could consider a bigger family of
deformations rather than simply shifts of the potential. In other
words, one could include any number of parameters on the same footing
as \vf\ in the analysis.  In the region of parameter space where
regular solutions are nongeneric, our method should apply.  We
consider a space in which one axis sets the initial condition
$\phi(0)$ and the other axes represent the values of the
parameters.  The number of passes is a well-defined function on this
space.  As above, regular solutions generically appear at the
boundaries where the number of passes changes.

\subsection{Differences between the diagrams}
\label{sec-differences}

Now let us discuss the differences between figures \ref{fig-tunnel}
and \ref{fig-notunnel}. When tunneling is forbidden at $V_{\rm F}=0$
(figure \ref{fig-notunnel}), nothing dramatic happens to the regular
single-pass instanton as $V_{\rm F} \to 0$. That it remains compact
may be seen by noting that at $V_{\rm F} =0$ there are two regular
single pass solution points, one at positive $\phi$ and one at
negative $\phi$. These two points are really the same solution; either
pole of the sphere can be considered the starting point. For $V_{\rm
  F} \leq 0$ our regular solutions persist, but since they are compact
they do not describe tunneling.

In contrast, figure \ref{fig-tunnel} describes a situation where
tunneling is allowed at $V_{\rm F}=0$.  This figure may look more
fine-tuned than the one which does not allow tunneling, because
various lines meet right at $V_{\rm F}=0$.  However, this behavior is
required, based on the results of the previous section showing that
the regular instanton is getting very large as $V_{\rm F} \to 0$. The
false vacuum pole of the one-pass instanton (on the left of the
diagram) gets closer and closer to the extremum.  This allows the
field to remain near the false vacuum for a very long time, resulting
in a big instanton that looks mostly like the background false vacuum
de~Sitter solution. (This behavior results in the singular one-pass
region on the left side of the diagram getting squeezed away and
disappearing in the limit.)

Also, on the right side of figure \ref{fig-tunnel}, the regular one-
and two-pass solutions must approach each other as $V_{\rm F} \to 0$,
squeezing the singular two-pass region. The two-pass
instanton~\cite{BouLin98} is a Euclidean solution which is symmetric
about the equator, so $\fd= 0$ at the equator.  For very large
instantons, two-pass solutions require $\fd$ to be zero at the
equator, while one-pass solutions require $\fd$ to be miniscule at the
equator so that it will be zero at the opposite pole. As $V_{\rm F}
\to 0$ and the instanton gets very large, the distinction between
these two conditions goes away and the instantons merge, becoming the
single-pass noncompact instanton for $V_{\rm F} \leq 0$.

Our diagrammatic method demonstrates other surprising properties of
the Euclidean solutions. For example, when tunneling is allowed, then a
solution which starts extremely close to the false vacuum will pass
the origin twice before escaping to infinity, while if tunneling is
not allowed it will only pass the origin once, and escape to the
opposite side. There is no obvious reason this property is related to
tunneling, but our diagram suggests it is the case and we have
verified this numerically for some potentials.

\subsection{Starting close to $\phi = 0$}
\label{sec-hm}

Let us explain another notable feature of our figures: The number of
oscillations at small amplitude decreases as the cosmological constant
increases.

For any value of the parameters, there is a trivial solution where the
field just sits at the local minimum of the Euclidean potential, the
Hawking-Moss instanton. Since we assume that this point has positive
vacuum energy ($U(0)<0$), the geometry is a four-sphere.  For sufficiently small
perturbations about this instanton, we can compute analytically the
number of passes.  Because the field probes only the immediate
neighborhood of its minimum, the Euclidean potential can be
approximated as a simple harmonic oscillator in this limit. Also, the
geometry can be taken to be just the Hawking-Moss instanton, since
corrections to the geometry due to oscillations of the field are
quadratic in the amplitude of the oscillations.

The Hawking-Moss instanton is
\begin{equation}
ds^2 = dt^2 + \rho(t)^2 d\Omega_3^2
\end{equation}
with
\begin{equation}
\rho(t) = \frac{1}{H_0} \sin(H_0t)~.
\end{equation}
Here $H_0$ is the Hubble constant determined by the value of the
potential at the Hawking-Moss instanton, 
\be
H_0^2 = \frac{8 \pi}{3M_{\rm Pl}^{2}} V(0)~.  
\ee
Small oscillations are then governed by the equation
\begin{equation}
\ddot{\phi} + 3 H_0 \cot(H_0 t) \fd = - U''(0) \phi ~.
\end{equation}
Defining the dimensionless variable $\tilde{t} = H_0 t$, the equation
takes the form of the eigenvalue equation for the Laplacian on a
$4$-sphere,\footnote{Once again, we thank Matt Kleban for pointing
  this out.}
\begin{equation}
\ddot{\phi} + 3 \cot(\tilde{ t}) \fd = -  {U''(0) \over H_0^2} \phi,
\end{equation}
where $\fd$ is now the derivative with respect to $\tilde{t}$. The
quantity $U''(0)/H_0^2$ plays the role of the eigenvalue.

Regular solutions exist only if the eigenvalue is of the form $n
(n+3)$ for a positive integer $n$.  These solutions are spherical
harmonics which are homogeneous on the three-sphere. They have $n$
zeroes, so for our purposes they are $n$-pass solutions.  Of course
the approximation we are using is only valid for infinitesimal
oscillations, so the physical statement is that as we shift the
potential, infinitesimal-amplitude regular solutions only exist
at special values.  This is consistent with our general rule that
regular solutions are nongeneric.

The number of passes will change as we shift the potential; $H_0$
changes if we add a constant $V_{\rm F}$ to the potential, while
$U''(0)$ is unaffected.  Recall that we denote the vertical shift of
the potential by \vf, so even though these solutions do not care about
the value of the energy in the false vacuum, in our notation we say
that infinitesimal-amplitude solutions exist at special values of \vf.
Between values of \vf\ allowing a regular solution with $n$ passes and
a regular solution with $n+1$ passes, the singular solutions have
$n+1$ passes.  So for singular solutions the number of passes is given
by the smallest integer $n$ such that
\begin{equation}
 n  (n+3) > {U''(0) \over H_0^2}~.
\label{maxno}
\end{equation}

For sufficiently large positive $V_{\rm F}$, it is clear from the
formula that the number of passes will be $n=1$. This happens because
the Hubble time for the Hawking-Moss instanton becomes short compared
to the period of the harmonic oscillator.

Once again, we see that regular solutions exist at the boundary
between singular solutions with different numbers of passes.  As
explained by Gratton and Turok~\cite{GraTur00}, the number of passes
is equal to the number of negative modes for perturbations around the
Hawking-Moss solution.  (The number of negative modes is important for
the interpretation of the instanton, but in this paper we are mainly
interested in the structure of the solution space.)

\subsection{Behavior as the maximum of the potential approaches $V =
  0$}
\label{sec-pile}

So far, we have assumed that the maximum of the potential 
(the Hawking-Moss point) has positive energy: $V(0)>0$. What happens if we allow $V(0) = 0$?  This corresponds to
the local minimum of the Euclidean potential moving up to $U=0$. The
results are dramatic. We can no longer argue that solutions are
generically compact, or generically singular, because there is a
new possible asymptotic behavior: the field can asymptotically
approach $\phi = 0$ while the geometry continues to grow ($\rho \to
\infty$).  This is possible because the potential is now zero at the
minimum, so asymptotically flat solutions are possible.

Furthermore, this asymptotic behavior is stable under small
perturbations, since the field approaches a local minimum of the
Euclidean potential.  So generic initial conditions can
result in noncompact, regular solution.  Specifically, all starting points
with Euclidean potential energy less than $U_{\rm F}$, the Euclidean
potential of the false vacuum, lead to this asymptotically flat
behavior; depending on parameters, a somewhat bigger set of starting
points leads to asymptotically flat solutions.  In fact, for the
potentials we will consider numerically $any$ starting point between $\phi_{\rm F}$ and $\phi_{\rm T}$ will lead to a noncompact solution with $\phi \to 0$.

To see that this is reasonable, recall the FRW equation
 \be H^2 = \frac{8
  \pi}{3 M_{\rm Pl}^{2}} \left[\frac{\dot{\phi}^2}{2}+U(\phi)\right] + {1
  \over \rho^2}~.  
 \ee
    As long as the field does not escape the region
of interest, $H$ cannot go to zero except at $\rho = \infty$ because
the Euclidean potential $U$ is nonnegative in this region. If $H$
cannot go to zero then it cannot become negative, so no anti-friction
is available. So solutions starting with energy less than $U_{\rm F}$ have
no hope of escaping because no anti-friction is available until after
they escape, and they would need to gain energy in order to escape.
Solutions beginning near the true vacuum, with more potential energy
than $U_{\rm F}$, can hope to overshoot on the first pass, but whether this
is possible depends on parameters.

There must be a dramatic signal of this lurking noncompactness as the
maximum of the potential approaches zero from above.  One place we can
see this is in our formula for small oscillations around the
Hawking-Moss solution.  Note that in this limit the Hubble constant
$H_0$ of the Hawking-Moss instanton approaches zero because the
cosmological constant of the Hawking-Moss solution is approaching
zero.  According to our formula Eq.~(\ref{maxno}) the number of passes
approaches infinity as $H_0 \to 0$.  The reason is that as $H_0 \to 0$
the Hubble time goes to infinity while the period of the harmonic
oscillator stays fixed, so there is time for an infinite number of
oscillations.

We have thus discovered an accumulation point, with
an infinite number of regular solutions appearing at $H_0 \to 0$. This
is shown schematically in figures~\ref{fig-notunnel}
and~\ref{fig-tunnel}, and will also appear in our numerical data. 

\subsection{Starting close to the vacua}
\label{subsec-startingclose}

Here we explain the behavior seen in both 
Fig.~\ref{fig-tunnel} and  Fig.~\ref{fig-notunnel} 
for starting points very close to the false vacuum. 
We will also be able to explain some aspects of the behavior just to the left
of the regular single-pass instanton and just to the right of the
true vacuum.
Note in the diagrams
that the number of passes is one for solutions whose starting point is 
just to the right of the false vacuum as long as $\vf > 0$.
However, as \vf\ becomes more negative the number of passes increases.

It is clear that the number of passes must be one for \vf\ positive.
The reason is that there is a regular compact solution with
the field sitting on the false vacuum- just the usual false vacuum
de Sitter solution. Also, starting points to the left of the false
vacuum must lead to zero passes by our assumptions about the
potential. Since the number of passes must change by one across 
a compact regular solution, starting points just to the right
of the false vacuum must lead to one-pass solutions.

This logic no longer holds once $\vf \leq 0$. 
There is still a regular solution where the field sits on top of the
false vacuum forever, represented by the vertical line in the
diagram, but this solution is now noncompact. We have no general
rules about how the number of passes can change across
a noncompact regular solution.

To see that starting points just to the right of the false vacuum
 might result in a large number of passes for \vf\ negative,
imagine a situation where \vf\ is very negative so that
both vacua are at very negative cosmological constant. (Note
that we are no longer interested in the limit of the previous section
where the Hawking-Moss solution becomes noncompact; here we
assume that it is compact.) If the field starts very close to the
false vacuum, it will sit there a long time before beginning to
 oscillate,
so that it essentially begins to oscillate at large $\rho$ compared to all other scales.

Recalling equation (\ref{eq-hdot}), the change in the Hubble parameter
is given by 
\be
\label{deltaH}
\Delta H = -\int ( 4 \pi M_{\rm Pl}^{-2} \fd^2 + 1/\rho^2)dt~.
\ee
We ignore the second term because we are at large $\rho$.
 The change in $H$ over one oscillation, given by the above formula,
is determined by the potential. It is basically constant as we
shift the potential vertically.
Now before the field starts oscillating the Hubble parameter is
positive, and its magnitude is
determined by the cosmological constant at the false vacuum, so its
magnitude is large for $\vf \ll 0$.
As a result, one oscillation of the field
produces a small change in $\Delta H/H$.

We have shown in section \ref{subsec-com} that any solution, singular or nonsingular, starting 
near the false vacuum must be compact.
 Since the solution must be compact, $H$ must become negative.
But since one oscillation leads to a small change in $\Delta H/H$,
it will take many oscillations before $H$ can become negative.
Further, the field experiences no anti-friction as long as
$H \geq 0$, so it is guaranteed to keep oscillating until
$H$ becomes negative.

Once $H$ becomes negative, more oscillations are necessary before
the field can escape (recall that generically the field escapes to
$\pm \infty$), 
because the magnitude of $H$ is bounded below by the vacuum
energy,
\be
H^2 \geq \frac{8 \pi}{3 M_{\rm Pl}^{2}}  U(\phi)~.
\ee
Another way to think about it is that the field has lost energy
to Hubble friction during its oscillations, and it must undergo
more oscillations during the phase of anti-friction to recover
enough energy to approach one of the vacua. These two points of
view are related by the FRW equation in the limit $\rho \to \infty$,
\be
H^2 = \frac{8 \pi}{3M_{\rm Pl}^{2}} E ~,
\ee
where $E = {1 \over 2} \dot{\phi}^2 + U(\phi)$
 is the Euclidean energy of the field.

Generically, the field will not return precisely to one of the
vacua after these oscillations; it will escape with the radius
is still extremely large.

To summarize, for starting points just to the right of the false
vacuum,
the generic solution consists of a very large region where the field
is essentially in the false vacuum. When the field begins
to oscillate there may be a large number of oscillations at large
$\rho$, after which the field escapes and the radius 
returns to zero. The field generically escapes at large radius.

Since there is a separation in scales between the size of the false
vacuum region and the characteristic period of oscillation, these
solutions should be well captured by the thin wall approximation.
The thin wall approximation will allow us to make more concrete
statements about the number of passes.\footnote{We 
emphasize that nowhere else in this paper
  do we appeal to a thin wall limit.}  
In order to capture the potentially large number of oscillations,
we slightly generalize the thin wall approximation: we allow
ourselves to stack multiple domain walls on top of each other,
each domain wall representing one pass.

We want to construct, within the thin wall approximation, a solution
(singular or nonsingular)
which has the features described above. It should 
have a regular pole surrounded by an enormous false vacuum
region surrounded by a stack of domain walls at infinity. 
On the other side
of the stack of domain walls, the radius should decrease back
to zero.
 We want to allow solutions with a singularity at one pole as
usual. Within the thin wall approximation, the only possible
singularity is a conical deficit. With a conical deficit, the metric
becomes
\begin{equation}
ds^2 = dt^2 + f^2 \rho(t)^2 d\Omega_3^2~,
\end{equation}
where $\rho(t)$ takes the usual Euclidean AdS form, $\rho(t) = R \sinh
(t/R)$, and $f>0$.  (For $f=1$ there is no conical deficit or excess.)

To summarize, we are seeking a thin wall solution which has a regular pole
surrounded by an enormous false vacuum region. The false vacuum region
is surrounded by a stack of domain walls. On the other side of the
domain
walls, we can have a region of true or false vacuum which generically
has a conical deficit. We expect that the thin-wall solution is a good
representation of the full solution in the false vacuum; we also expect
it to capture correctly the number of field oscillations required before
the field has enough energy to escape. However, as mentioned above
the field generically escapes at large radius, and after this point it
is not near either of the vacua, so the thin wall approximation must be
wrong. Nevertheless, the thin wall approximation should be a good
one for predicting the  number of oscillations.

The Israel junction condition relates the jump in extrinsic curvature
across the stack of domain walls to the tension. For the situation we are
considering, in which the radius decreases as we move away from the
domain wall in either direction, it is
 \begin{equation}
 \label{junction}
   \sqrt{{1 \over  (f_1 \rho)^2} + {1 \over R_1^2}} 
   + \sqrt{{1 \over (f_2 \rho)^2} +  {1 \over R_2^2}}  = n \sigma~,
\end{equation}
where $R_1$ and $R_2$ are the AdS radii on either side of the wall,
$f_1$ and $f_2$ are related to the conical deficits, $\sigma$ is
proportional to the tension of one domain wall, and $n$ is the
number of domain walls in the stack. The left side of the equation is bounded
below by $1/R_1 + 1/R_2$~. Since the first pole is nonsingular by
construction, we set $f_1 = 1$.

\subsubsection{Vanishing false vacuum energy}

We first consider the special case where the false vacuum has zero
cosmological constant, $R_{\rm F} = \infty$. Above,
we claimed that if the false vacuum has very negative cosmological
constant, the number of passes (the number of domain walls
in the thin wall approximation) will be large. On
the other hand, if
the false vacuum has zero cosmological constant we will see that
one or two domain walls is sufficient.
The simplest solution would be
a large region of false vacuum, 
one  domain wall at large radius, and a region of true vacuum
on the other side. In this case the left side of Eq.~(\ref{junction}) is bounded below by $1/R_{\rm T}$.
If the tension is big enough,  $\sigma > 1/R_{\rm T}$,
a solution exists. In this case, for sufficiently large $\rho$, 
we can always solve the junction
condition for the conical deficit parameter $f$.

On the other hand, if $\sigma \leq 1/R_{\rm T}$, no solution exists.
In this case, we will need two domain walls to solve the junction
condition. With an even number of domain walls we have false
vacuum on both sides. The junction condition becomes
\be {1 \over \rho} + {1 \over f \rho} = 2
\sigma \label{eq-frho} \ee 
  This equation
can be solved for any tension $\sigma$, because
the left side can take any value. For fixed sufficiently
large $\rho$ we can again solve for $f$.

So we need one domain wall if $\sigma > 1/R_{\rm T}$
and two domain walls if $\sigma < 1/R_{\rm T}$
But $\sigma < 1/R_{\rm T}$ is precisely the condition that
the false vacuum can decay! Equivalently, it is the
condition that a noncompact regular instanton exists.
The conclusion is that if tunneling is forbidden
(Fig. \ref{fig-notunnel}), 
starting points
just to the right of the false vacuum have one pass,
while if tunneling is allowed (Fig. \ref{fig-tunnel}) 
they  have two passes.

 Another way to understand that two domain walls are required
when tunneling is allowed is to note that 
Euclidean AdS space has finite extrinsic curvature at
infinity.  Hence, joining an AdS region to a flat region across a three-sphere
requires a jump in the extrinsic curvature that remains finite even as
the location of the domain wall $\rho$ goes to infinity.  If tunneling
is allowed, the tension is too small to account for this finite
change.

Now let us move to the right part of Fig.~\ref{fig-tunnel}.  We can
infer the behavior of solutions which start just to the left of the
single-pass noncompact instanton, that is, solutions which barely
undershoot the false vacuum. The regular single-pass instanton
achieves the false vacuum at very large $\rho$; a slightly undershooting
solution will spend a long time close to the false vacuum before
falling back. Since we have just analyzed solutions which spend a long
time and grow to large radius near the false vacuum, the subsequent
evolution will be the same. 

In the thin wall approximation, the equivalent statement is that the
single-pass instanton and the pure false vacuum solution have the same
asymptotics, so we can replace the false vacuum region in our previous
solution by the single-pass regular instanton.  The conclusion is that
a solution starting just to the left of the regular instanton will
have one more pass than a solution which starts just to the right of
the false vacuum. This explains why the number of passes jumps from
one to three across the regular instanton.

\subsubsection{Negative false vacuum energy}

Next, let us
consider the lower part of Fig.~\ref{fig-tunnel} and  Fig.~\ref{fig-notunnel},
the region with $\vf < 0$. Again we folcus on starting points
just to the right of the false vacuum. Now it is no
longer clear that we can have a solution with two domain walls,
because even the false vacuum has finite extrinsic curvature as $\rho
\to \infty$; as we argued below Eq.~(\ref{deltaH}) a large number of domain walls may be necessary.  Two types of solutions are possible: with an even number
of domain walls, we have false vacuum on both sides, and for a
solution to exist we need 
\be 1/R_{\rm F} + 1/R_{\rm T} < n \sigma, \ \ \ \ n \
{\rm even.}
\label{even}
\ee For an odd number of domain walls, we have true vacuum on one
side, and for a solution to exist we need \be 1/R_{\rm F} + 1/R_{\rm T} < n
\sigma, \ \ \ \ n \ {\rm odd.}
\label{odd}
\ee

The minimum number of passes is determined by the smallest number $n$
of domain walls such that one of the inequalities is satisfied. It is
clear that as \vf becomes more negative the required number of passes
$n$ becomes larger.  This is reflected in both diagrams.

Additionally, if tunneling is allowed for a given $V_{\rm F}\leq 0$,
then the minimum number of passes always occurs for $n$ even. This is
because the condition that tunneling is allowed is $1/R_{\rm T} - 1/R_{\rm F} <
\sigma$. When this is satisfied, equation (\ref{even}) is always
satisfied for smaller $n$ than equation (\ref{odd}). The result is that
along the left side of the diagram, all numbers should be even for
$\vf \leq 0$ as long as tunneling is allowed.

Let us move to the right of the figures.  As in the $V_{\rm F}=0$
case, we parlay our results about solutions starting near the false
vacuum into information about starting points just to the left
 of the single-pass
regular instanton.  Once again the
number of passes just to the left of the noncompact instanton
should be one greater than the number of passes for a starting point
just to the right of the false vacuum for fixed \vf.

Hence, if tunneling is allowed, only odd numbers
should appear immediately to the left of the regular instanton. On the other hand, if tunneling is forbidden then as we dial $\vf$ the
number of passes $n$ near the false vacuum will change by one at a
time, though there will be much more parameter space where the number
is even.

\section{Numerical Evidence}
\label{sec-numerical}

\subsection{Method}
\label{subsec-method}

In order to exhibit the features discussed in the previous sections,
we turn to numerical solutions of the CDL instanton equations
(\ref{unscaledCdL}).  

Following Ref.~\cite{BJ}, we rescale the physical quantities to yield
dimensionless variables.  The potential for $\phi$ can be written
\begin{equation}
V(\phi) = \mu^4 v(x)~,
\end{equation} 
where
\begin{equation}
x=\phi/M~.
\end{equation} 
Here, $M$ is the scale over which the potential has nontrivial
features, and $\mu^4$ is a characteristic energy density.  We assume
that $\mu\ll M\leq M_{\rm Pl}$ where $M_{\rm Pl}$ is the Planck scale.
The two mass scales have been extracted so that $v$ is a function that
can be approximated as a polynomial with coefficients of order one.

\begin{figure}
\begin{center}
\includegraphics{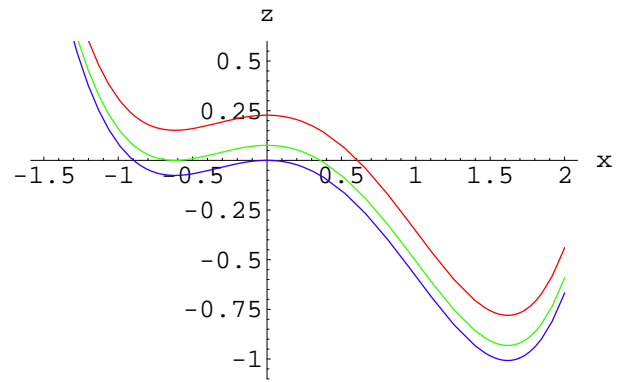}
\end{center}
\caption{A graph of the potential $v(x)$ for $b=1$, and $z = \{1,0,-1\}$.} 
\label{fig-potential1}
\end{figure} 

\begin{figure}
\begin{center}
\includegraphics{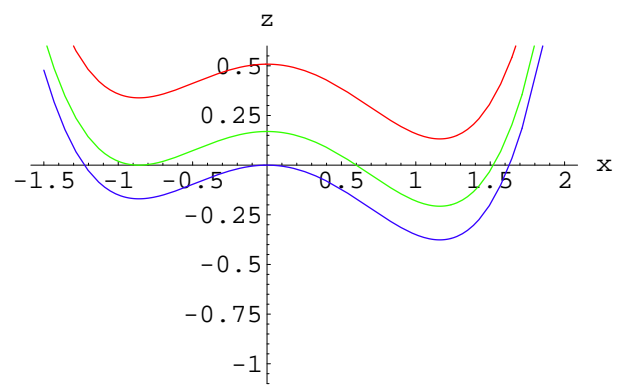}
\end{center}
\caption{A graph of the potential $v(x)$ for $b=0.3$, and $z = \{1,0,-1\}$. }
\label{fig-potential3}
\end{figure} 

We employ the same quartic potential used in Ref.~\cite{BJ}, which
takes the form
\begin{equation}
\label{potential}
v(x) = f(x) - (1+z) f(x_{\rm F})~,
\end{equation}
where
\begin{equation}
\label{f}
f(x) = \frac{x^4}{4} - \frac{b x^3}{3} - \frac{x^2}{2}~.
\end{equation}
The function $f(x)$ has two negative local minima at $x_{\rm F}$ and
$x_{\rm T}$, with $f(x_{\rm F})>f(x_{\rm T})$, and a local maximum at
$x=0$.  The potential $v(x)$ share these properties. 

The adjustable parameter $z$ allows us to tune the false vacuum energy
by shifting the entire potential with $v(x_{\rm F}) \propto z$.  Thus,
$z$ plays the role of the shift $V_{\rm F}$ in the previous sections.
In particular, the interesting limit $V_{\rm F}\to 0$ corresponds to
$z\to 0$.  (Note that $f(x_{\rm F})<0$, so $z>0$ corresponds to
$V_{\rm F} >0$.)

The parameter $b$, which we take to be strictly positive, controls
both the width $x_{\rm T} - x_{\rm F}$ and relative heights of the vacua $v(x_{\rm
  F}) - v(x_{\rm T})$.  Figures \ref{fig-potential1} and
\ref{fig-potential3} show plots of $v(x)$ of $b = 1$ and $0.3$,
respectively, for various values of $z$.

The radius and time can also be made dimensionless by rescaling by
appropriate powers of the mass scales $\mu$ and $M$:
\begin{eqnarray}
\label{dimlessvars}
r &=& \frac{\mu^2 \rho}{M}~, \\
s &=& \frac{\mu^2 t}{M}~, \\
\epsilon &=& \frac{M}{\sqrt{3} M_{\rm Pl}}~.
\end{eqnarray}
We have also defined a dimensionless quantity $\epsilon$ controlling
the strength of gravity.  For $\epsilon\ll 1$, gravity has a
negligible effect on the decay rate, but when $\epsilon$  is of order one,
gravity can be important.  For example, it can completely suppress the
decay of flat space.

Let us rewrite the Euclidean CDL equations (\ref{unscaledCdL}) and
(\ref{eq-constr}) in terms of these dimensionless variables:
\begin{eqnarray} 
\label{scaledCdL}
&&\ddot{r} = -\epsilon^2 r \left[\dot{x}^2 + v(x) \right]~, \\
&&\ddot{x} + 3 \frac{\dot{r}}{r} \dot{x} = v'(x)~, \\
&&\dot{r}^2-1 = \epsilon^2 r^2 \left[\frac{\dot{x}^2}{2}-v(x)\right]~.
\end{eqnarray}
where an overdot (prime) denotes differentiation with respect to $s$
($x$). 

Using Mathematica
, we numerically integrate these equations with initial boundary
conditions
\begin{eqnarray}
\label{boundcond}
\dot{r}(0) = 1 &,& r(0) = 0~, \\
\dot{x}(0) = 0 &,& x(0) = x_0
\end{eqnarray}
to yield functions $r(s)$ and $x(s)$.  The initial position $x_0$
along with $z$, $b$, and $\epsilon$ form a set of four adjustable
parameters on which the solutions depend.

By numerically constructing many solutions, both singular and regular,
for a variety of parameters, we will be able to verify the assertions
made in sections \ref{sec-cont} and \ref{sec-structure}.  For generic
choices of parameters, the numerical integration produces compact
singular solutions where the field $x(s)$ inevitably escapes to $\pm
\infty$, as explained in Sec.~\ref{sec-escape}.  For regular compact
solutions, the unbounded anti-friction as $r \to 0$ makes the numerics
difficult to control near the far pole.  By tuning $x_0$, however,
regular solutions can be very well approximated. The range of $r$ over
which $x(s)$ can be made to remain near $x_{\rm T}$ or $x_{\rm F}$
before rolling off to $\pm \infty$ is limited only by the tuning of
$x_0$ and calculational precision.

Noncompact solutions can likewise be approximated by fine-tuning
$x_0$.  In accordance with the arguments of section \ref{sec-cont},
these are necessarily regular one-pass solutions starting at the true
vacuum and asymptoting to the false vacuum.  As with the approximately
regular compact solutions, the inevitable result of imperfect tuning
of $x_0$ is that $x(s)$ eventually either over- or under-shoots
$x_{\rm F}$, leading to a singularity.  However, with sufficient
tuning, we can obtain a large (in $s$) region where $\dot{r}=1$ and
reach a large maximum $r$ before the singularity.

\subsection{The decay of nearly flat space}
\label{subsec-nearlyflatdecay}


Our first goal is to confirm the result of Sec.~\ref{sec-cont} that
the decay rate is continuous as $z\to 0$. 

As was pointed out in Ref.~\cite{CDL}, for certain potentials a
seemingly metastable Minkowski space ($z=0$) can be stabilized by
gravitational effects.  To exhibit this effect, we tune $\epsilon$
which controls the strength of gravitational corrections. Keeping
$b=1$ fixed, we find a critical value of $\epsilon \approx .35$.
Below this value, the $z=0$ one-pass solution is noncompact, implying
the instanton corresponds to an actual decay of Minkowski space.
Above the critical value, the regular one-pass solution is compact at
$z=0$ and therefore does not represent decay.  A similar transition
occurs when varying $b$ for fixed $\epsilon$. For fixed $\epsilon =
0.6$, for example, the $z=0$ regular one-pass solutions are compact
below $b \approx 0.74$ and noncompact above this value.

We now turn to the question discussed in Sec.~\ref{sec-cont}: the
behavior of compact, regular one-pass $z > 0$ solutions in the limit
where $z \to 0$.  The asserted continuity of the decay rate translates
into different behavior, depending on whether the false vacuuum at $z
= 0$ is stable or not.  We chose $\epsilon=0.6$, $b=.3$ to study the
stable case and $\epsilon=0.6$, $b=1$ for the unstable case.

\begin{figure}
\begin{center}
\includegraphics{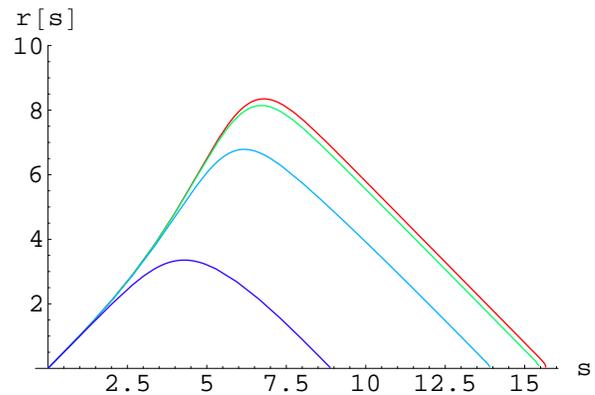}
\end{center}
\caption{This graph shows a family of one-pass solutions with $b=0.3$,
  $\epsilon = 0.6$, and $z = \{1, 0.1, 0.01, 0 \}$.  As
  $z \to 0$, these compact solutions smoothly approach the $z=0$
  solution, which for this value of $(b,\epsilon)$ is also compact.
  The decay of flat space is forbidden.  This reproduces the results of \cite{BJ}}
\label{fig-compact}
\end{figure} 

For each case, we probed the $z \to 0$ limit by numerically computing
a family of regular single-pass solutions with $z = \{1, 0.1, 0.01, 0
\}$.  The stable case is shown in figure \ref{fig-compact}.  As
expected, the family of compact solutions smoothly approaches the
compact $z=0$ solution.  This is the behavior anticipated in
Sec.~\ref{sec-cont}.

\begin{figure}
\begin{center}
\includegraphics{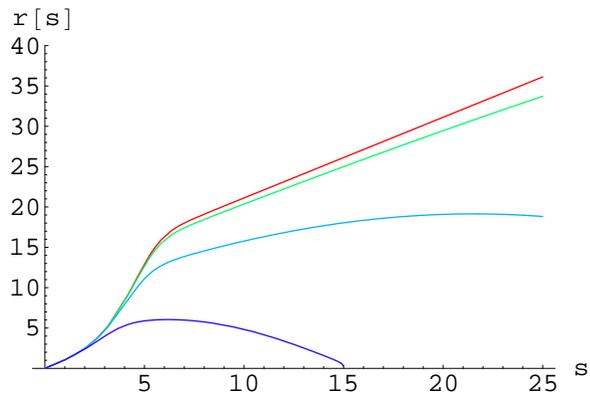}
\end{center}
\caption{This graph shows a family of one-pass solutions with $b=1$,
  $\epsilon = 0.6$, and $z = \{1, 0.1, 0.01, 0 \}$.  As $z \to 0$,
  these compact solutions smoothly approach a noncompact $z=0$
  solution.  Flat space decays.}
\label{fig-noncompact}
\end{figure} 

In the unstable case, the compact $z > 0$ solutions dramatically grow
in size as $z \to 0$, as shown in figure \ref{fig-noncompact}.  For $z
\le 0.1$ we lack the numerical precision to follow the solution all
the way to the far pole at $s > 30$ because the corresponding dS
spheres are growing so large.  For $z = 0.1$ at least the equator
where $\dot{r} = 0$ at $s = 22$ is still within our computational
range.  We can clearly infer, though, that as $z \to 0$ the compact
instantons are growing so as to reach infinite size in the limit.  The
$z= 0.01$ and $z = 0$ solutions are virtually indistinguishable.
Again, this supports the behavior anticipated in Sec.~\ref{sec-cont}.

In summary, our analytic argument that the decay rate is continuous is
borne out by the numerical evidence.

\subsection{Plots of the solution space}
\label{subsec-plots}

Our next numerical goal is to verify the broader analysis of the CDL
solution space provided in section \ref{sec-structure}.  For fixed
$\epsilon = 0.6$, we considered the two important cases $b=1$ (flat
space decays) and $0.3$ (flat space is stable).
For each case, we computed the number of passes for $10^6$ solutions
with $z \in [-1,9]$ and $x_0 \in [x_{\rm F}, x_{\rm T}]$.  The
resolution is approximately $\delta x_0 \sim 10^{-3}$ and $\delta z
\sim 10^{-2}$. 

\begin{figure}
\begin{center}
\includegraphics{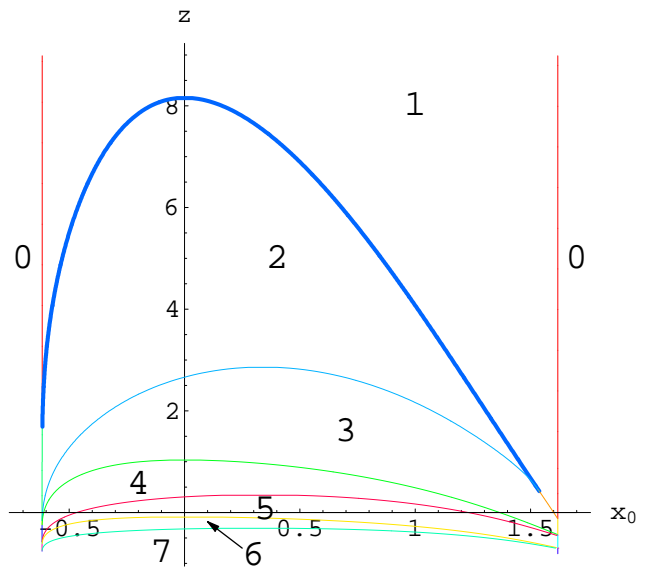}
\end{center}
\caption{The $(x_0, z)$ solution space for $b=1$ and $\epsilon = 0.6$.
  This numerical plot reproduces the general features of figure
  \ref{fig-tunnel}.  {\em Here, tunneling is allowed at $z=0$.}  The
  numbers $0$ to $7$ denote the number of passes for solutions in a
  given region. The curves at the boundaries between regions are
  regular solutions.  The vertical lines are the trivial solutions at
  $x_{\rm T}$ and $x_{\rm F}$, and the thick blue line is the
  single-pass instanton.  The Hawking-Moss solution (a vertical line
  at $x=0$) is hidden by the axis.  Features near $x_{\rm T}$ and $x_{\rm F}$ are not well-resolved; see figures \ref{fig-ultrafineleft1} and \ref{fig-ultrafineright1}}
\label{fig-rainbow1}
\end{figure}

\begin{figure}
\begin{center}
\includegraphics{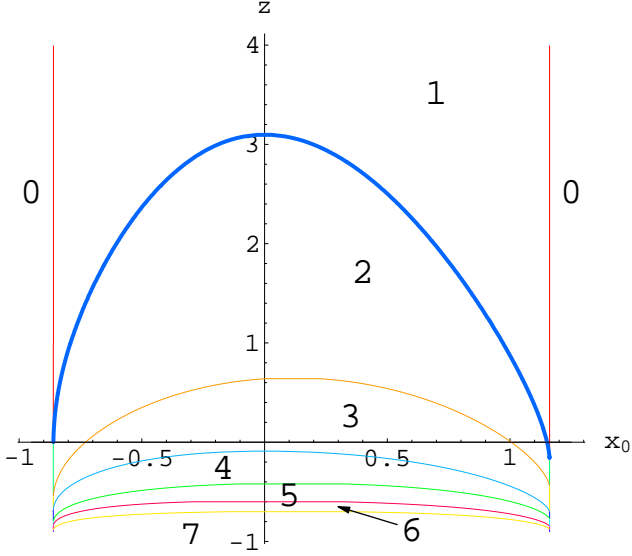}
\end{center}
\caption{The $(x_0, z)$ solution space for $b=0.3$  and
  $\epsilon = 0.6$.  This numerical plot reproduces the general
  features of figure \ref{fig-notunnel}.  {\em Tunneling is not
    allowed at $z=0$.}  The numbers $0$ to $7$ denote the number of
  passes for solutions in a given region, and the curves represent
  both regular solutions and the boundaries between regions.  The
  vertical lines are the trivial solutions at $x_{\rm T}$ and $x_{\rm
    F}$, and the thick blue line is the single-pass instanton.  Features near $x_{\rm T}$ and $x_{\rm F}$ are not well-resolved; see figures \ref{fig-ultrafineleft3} and \ref{fig-ultrafineright3}}
\label{fig-rainbow3}
\end{figure}

The results are shown in figures \ref{fig-rainbow1} and
\ref{fig-rainbow3}.
The number of passes for each region is labeled.  The curves
demarcating the boundaries between regions with different numbers of
passes are the locations of the regular solutions.

\begin{figure}
\begin{center}
\includegraphics{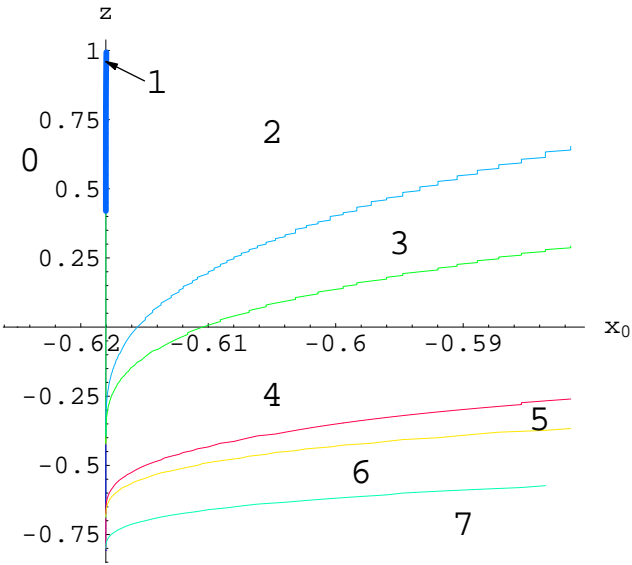}
\end{center}
\caption{A close-up of the lower-left corner fig.~\ref{fig-rainbow1}
  (tunneling allowed at $z=0$), showing the numerically computed
  $(x_0, z)$ solution space with $b=1$ and $\epsilon = 0.6$.  The
  thick, blue one-pass curve is nearly vertical and is merging with
  the $x_{\rm F}$ solution.  Although we expect that they meet at
  $z=0$, our limited precision makes them appear to merge at higher
  $z$.}
\label{fig-ultrafineleft1}
\end{figure}

\begin{figure}
\begin{center}
\includegraphics{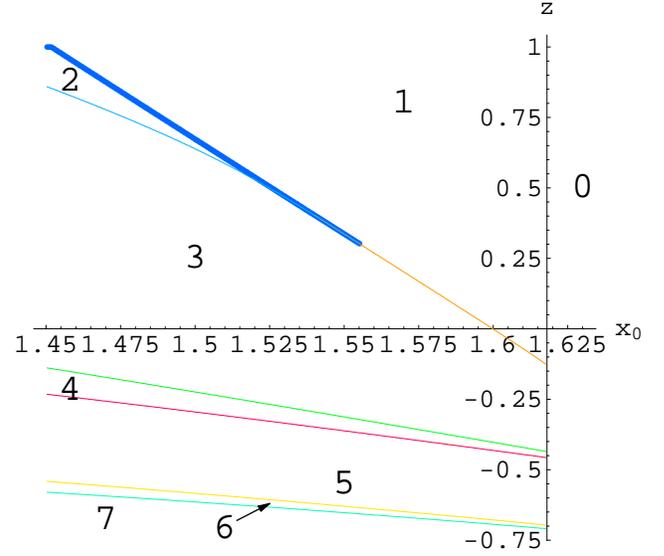}
\end{center}
\caption{A close-up of the lower right corner of
  fig.~\ref{fig-rainbow1} (tunneling allowed at $z=0$) showing the
  numerically computed $(x_0, z)$ solution space with $b=1$ and
  $\epsilon = 0.6$.  The regular one- (thick, blue line) and two-pass
  solutions merge as $z$ decreases.  Due to our limited precision,
  they appear to meet at $z = 0.3$ rather than at $z=0$ as expected.}
\label{fig-ultrafineright1}
\end{figure}

\begin{figure}
\begin{center}
\includegraphics{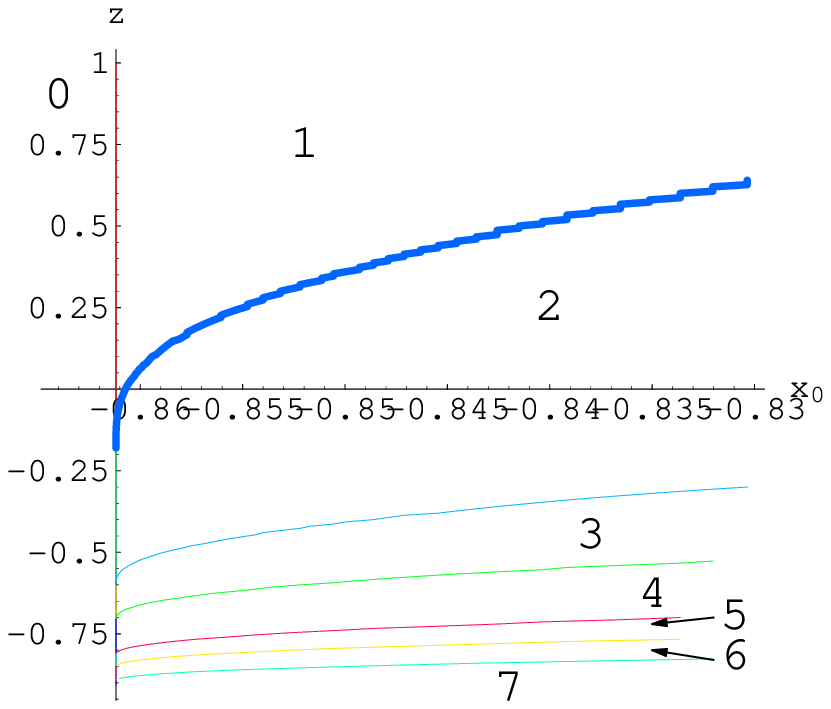}
\end{center}
\caption{A close-up of the lower left corner fig.~\ref{fig-rainbow3}
  (tunneling forbidden at $z=0$) showing the numerically computed
  $(x_0, z)$ solution space with $b=0.3$ and $\epsilon = 0.6$.  One
  can see that the one-pass regular solution passes through $z=0$
  without merging.}
\label{fig-ultrafineleft3}
\end{figure}

\begin{figure}
\begin{center}
\includegraphics{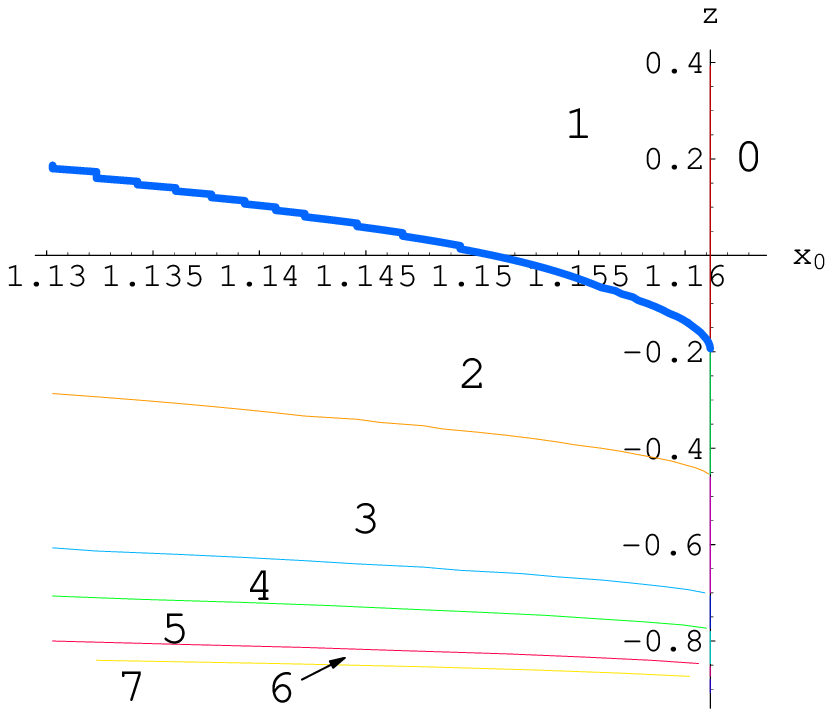}
\end{center}
\caption{A close-up of the lower right corner fig.~\ref{fig-rainbow3}
  (tunneling forbidden at $z=0$) showing the numerically computed
  $(x_0, z)$ solution space with $b=0.3$ and $\epsilon = 0.6$.  Again,
  one can see that the one-pass regular solution passes through $z=0$
  without merging.}
\label{fig-ultrafineright3}
\end{figure}

To better access features near $x_{\rm F}$ and $x_{\rm T}$, we also
ran $10^5$ points close to each edge for each value of $b$, with
exponentially decreasing step size in $x_0$ as the edge was approached.
Our resolution was then $\delta \log(x_0) \sim 10^{-1}$ and $\delta z \sim 10^{-2}$.
The magnified left edges of figures \ref{fig-rainbow1} and
\ref{fig-rainbow3} are shown in figures \ref{fig-ultrafineleft1} and
\ref{fig-ultrafineleft3}.  Likewise, figures \ref{fig-ultrafineright1}
and \ref{fig-ultrafineright3} zoom in on the right edge.


\subsection{Discussion}
\label{subsec-disc}

A number of important features predicted in figures \ref{fig-tunnel}
and \ref{fig-notunnel} are reproduced numerically in figures
\ref{fig-rainbow1} and \ref{fig-rainbow3}.  As we argued in
Sec.~\ref{sec-jump}, when crossing any regular compact solution, the
number of passes should jump by one.  This can be seen in figures
\ref{fig-rainbow1} and \ref{fig-rainbow3}.  It can also be seen that curves
representing noncompact solutions separate singular regions whose
number of passes differ by as much as seven.

Another notable feature of the $z \to 0$ limit in the unstable case
(figure \ref{fig-tunnel}) was the merging of solutions as $z \to 0$.
As explained in section \ref{sec-differences}, the regular one-and two-pass
solutions on the right merge into each other at $z=0$, as do the zero-
and one-pass solutions on the left.  This was the most obvious feature
distinguishing the unstable from the stable diagram (figure
\ref{fig-notunnel}).  

This merger can be roughly made out in our numerical plot for the
unstable case, figure \ref{fig-rainbow1}; and it is notably absent in
the stable plot, figure \ref{fig-rainbow3}.  This is best seen in the
refined plots for the left and right edge; note how the thick blue
line passes through $z=0$ in figures \ref{fig-ultrafineleft3} and
\ref{fig-ultrafineright3}).  This supports our arguments in
Sec.~\ref{sec-differences}.

However, neither figure \ref{fig-rainbow1}, nor its refinements,
figures \ref{fig-ultrafineleft1} or \ref{fig-ultrafineright1}, 
are able to show that the merger takes place exactly at $z=0$.  The
curves become so close together that even at $z=0.1$, a
resolution of $\delta x_0 \sim 10^{-10}$ is insufficient to detect the
region between them.

The plots also confirm the estimate in Sec.~\ref{sec-hm} for the number
of oscillations for solutions near $x_0 = 0$: The regular solution
with the largest number of passes at a given $z > -1$ is given by the
largest integer $n$ satisfying Eq.~(\ref{maxno}), which in
dimensionless variables is
\begin{equation}
\label{maxnodimensionless}
n(n+3) < \frac{|v''(0)|}{\epsilon^2 v(0)}~.
\end{equation}
For example, in the $b=1$ case, at $z=1$ the right-hand side of Eq.~(\ref{maxnodimensionless}) is $18.3$
which yields $n=3$.  From figure \ref{fig-rainbow1}, we can see the
curve of regular three-pass solutions crosses the origin just above
$z=1$. Below it, in particular at $z=1$ near $x=0$, are singular
three-pass solutions.

Confirming a prediction from Sec.~\ref{sec-pile}, the regular solution
curves begin to pile up when $z \to -1$, in both figure
\ref{fig-rainbow1} and figure \ref{fig-rainbow3}.  Ultimately, the
resolution is insufficient to detect every transition between
constant-pass regions.  As $z \to -1$ the number of oscillations
increases rapidly, the amplitude decreases, and the size of the
solutions grow very large, $s \gg 1$.  For example, when $b=1$ ,
$z=0.99$, and $|x_0 - x_{\rm T}| = 10^{-5}$, $x(s)$ oscillates for so
long around the Hawking-Moss solution, that the numerical precision is
insufficient to follow the solution back to $r=0$, and instead errors
build up and lead to a singularity at $s \sim 130$.

The plots also connect with the discussion in Sec.~\ref{sec-structure}.  For
$z<0$, the regular solutions appear to merge in pairs before reaching
$x_{\rm F}$ or $x_{\rm T}$.  However, we can resolve the two curves,
and, when we zoom in on the edges by plotting figures
\ref{fig-ultrafineleft1} and \ref{fig-ultrafineleft3} logarithmically
in $x_0 - x_{\rm F}$ in or figures \ref{fig-ultrafineright1} and
\ref{fig-ultrafineright3} logarithmically in $x_{\rm T} - x_0 $, we
can see the curves reach $x_{\rm F}$ or $x_{\rm T}$ at separate
points.

\acknowledgements

We would like to thank Anthony Aguirre, Tom Banks, and Matthew Johnson
for stimulating arguments.  We are especially grateful to Matthew
Kleban for suggesting the rainbow diagram.

\bibliographystyle{board}
\bibliography{all}

\end{document}